\documentclass{svproc}
\usepackage{url}

\usepackage{graphicx}
\usepackage{multicol}
\usepackage{footmisc}
\usepackage{gensymb}
\usepackage{wasysym}
\usepackage[table]{xcolor}

\begin{document}
\mainmatter              
\title{Insight into the origin of cometary ices from Rosetta/ROSINA mass spectrometer data}
\titlerunning{Origin of the ices in comet 67P/C-G}  
%
\author{Martin Rubin}
%
\authorrunning{Martin Rubin} 
%
\tocauthor{Martin Rubin}
\institute{Physikalisches Institut, University of Bern, Sidlerstrasse 5, CH-3012 Bern, Switzerland,\\
\email{martin.rubin@unibe.ch}
}

\maketitle              

\begin{abstract}
Here we review some of the major findings of the mass spectrometer suite ROSINA on board of ESA's Rosetta spacecraft to comet 67P/Churyumov-Gerasimenko. For more than 2 years, ROSINA continuously measured the composition of the gases sublimating from the comet's nucleus. ROSINA measurements provided insight into the origin of the ices in 67P/Churyumov-Gerasimenko. The obtained molecular, elemental, and isotope abundances revealed a composition more complex than previously known. Furthermore, a subset of these measurements indicate that a substantial fraction of the molecules incorporated into the comet predate the formation of the solar system. 


\end{abstract}
\section{The Rosetta mission}
The European Space Agency's Rosetta mission accompanied and thoroughly investigated comet 67P/Churyumov-Gerasimenko (hereafter 67P) for over two years, starting with the arrival at the comet in early August 2014 and concluding with the orbiter's final descent to the surface of the nucleus at the end of September 2016. 
Rosetta was sent out to investigate the nucleus interior and physical process on its surface, the dust and gas coma, and its interaction with the solar wind \cite{Glassmeier2007}. In particular, the science goals included ``the determination of chemical, mineralogical and isotopic compositions of volatiles and refractories in the cometary nucleus'' and ``the study of the development of cometary activity''.

\subsection{\label{sec:67P}Comet 67P/Churyumov-Gerasimenko}
Comet 67P is a Jupiter-family comet (JFC) which is on its current \mbox{6.4-year} elliptical orbit since a close encounter with Jupiter in 1959 \cite{Maquet2015}. 67P has a pronounced bi-lobate shape, assumed to be the result of a collisional merger \cite{Jutzi2015}, and the tilt of its rotation axis leads to pronounced seasonal outgassing \cite{Hassig2015}. During the two years Rosetta accompanied the comet, 67P lost about 1$\permil$ of its mass through outgassing and associated dust mass loss \cite{Patzold2019}. This resulted in erosion rates of up to several meters, especially on the southern hemisphere which experienced a short but intense summer around perihelion at 1.24~au.

\subsection{The ROSINA ion and neutral gas mass spectrometer suite}
Rosetta carried a complementary set of payload instruments, among them ROSINA, the Rosetta Orbiter Spectrometer for Ion and Neutral Analysis. ROSINA consisted of three instruments, the Double Focusing Mass Spectrometer DFMS, the Reflectron-type Time-Of-Flight mass spectrometer RTOF, and the COmet Pressure Sensor COPS \cite{Balsiger2007}. 
In RTOF, DFMS, and COPS the neutral gas entering the instrument was ionized by electron-impact. In DFMS, a set of electrostatic and magnetic fields were used to select ions with a given mass/charge before detection on a Micro Channel Plate (MCP) detector. In RTOF, the ions were extracted by a sharp pulse and their flight time to the MCP detector was measured and converted to mass/charge. In COPS, the current of newly formed ions was measured and relates to the local neutral gas density.

\section{ROSINA measurements pertaining to the origin of the material in comet 67P}
ROSINA was dedicated to address several science goals of the Rosetta mission with focus on the origin of the ices in 67P and the comet's interaction with the Sun \cite{Balsiger2007}. For this purpose DFMS, RTOF, and COPS measurements were used to determine the chemical and isotopic composition of the volatile material in the coma and to monitor the outgassing while the comet covered a range of heliocentric distances through the inner solar system. ROSINA was operated almost continuously throughout the Rosetta mission. The major gases in the coma of 67P were H$_2$O, CO$_2$, and CO, followed by a plethora of minor species \cite{Rubin2019a}, some of them measured for the first time in a comet. In the following, a subset of key measurements regarding the origin of the cometary material will be discussed.

\subsection{Molecular oxygen}
First, already early in the Rosetta mission a surprising amount of O$_2$ with a relative abundance on the order of (3.80$\pm$0.85)\% with respect to H$_2$O was found in 67P's coma \cite{Bieler2015a}. The ratios near perihelion and integrated over the whole mission were comparable, i.e. (3.1$\pm$1.1)\% \cite{Rubin2019a} and (2.3$\pm$0.5)\% \cite{Lauter2020}, respectively.  A re-analysis of Neutral Mass Spectrometer data from ESA's Giotto flyby at comet \mbox{1P/Halley} in 1986 indicated a similar O$_2$/H$_2$O ratio of (3.70$\pm$1.7)\% \cite{Rubin2015a}. This suggests that O$_2$ may be a rather common species in comets. Furthermore, measurements showed that O$_2$ was rather well correlated to H$_2$O, despite the very different sublimation temperatures \cite{Fray2009}. Several possible formation mechanisms of the O$_2$ were discussed, however, processes such as radiolysis of water ice \cite{Mousis2016a}, dismutation during water ice desorption \cite{Dulieu2017}, or Eley-Rideal reactions of energetic water ions with oxidized surfaces on the nucleus or dust grains \cite{Yao2017} seem to be at odds with the different oxygen isotope ratios found in both molecules (see Tab.~\ref{tab:isotopes}). Furthermore, meters of erosion per orbit \cite{Keller2015a} continuously expose fresh material from the comet's interior which limits processing of the material through, e.g., chemical reactions and irradiation. And also fluxes of water ions are too low to explain the observed amount of O$_2$ \cite{Heritier2018}. More promising is a presolar origin \cite{Altwegg2019}, possibly through gas-grain chemistry \cite{Taquet2016}.

\subsection{\label{sec:highlyvolatiles}Highly volatile species}
A suite of highly volatile species, including CO, N$_2$, CH$_4$, and O$_2$ \cite{Fray2009}, were present in the coma throughout the Rosetta mission. A JFC, like 67P, is expected to have undergone heating of up to 60~K in the top few 100~m during the several million years on a Centaur orbit at $\sim$7~au before becoming a JFC \cite{Guilbert2016}. This may lead to a gradual diffusion and loss of highly volatile species, the noble gas neon, e.g., was not detected \cite{Rubin2018}. Indeed, 67P seems to be depleted in highly volatile species compared to Oort cloud comets \cite{Altwegg2019}. On the other hand, erosion processes may again provide access to fresh material taking into account that the comet visited the inner solar system already several times \cite{Maquet2015}. Still, these measurements indicate that comets, such as 67P, were never part of a larger parent body and hence warm, e.g., due to radiogenic heating \cite{Mousis2017}. Additional evidence for 67P's cold origin and storage will be discussed later in section~\ref{sec:d2h}.

The situation, however, is further complicated given that some of these highly volatile species are present at the sub-percent level. As a result, their outgassing behavior is strongly influenced by the major volatiles H$_2$O and CO$_2$. While pure ices, made up of species such as CH$_4$ and N$_2$, may already be lost due to their high volatility, co-desorption of the fraction trapped in H$_2$O and CO$_2$ (or loss associated to phase transitions of, e.g., H$_2$O) is likely key in their release \cite{Kouchi1995}.

\subsection{\label{sec:chemicalcomplexity}Chemical complexity and comparison to ISM}
\begin{figure}
\includegraphics[width=1.0\textwidth]{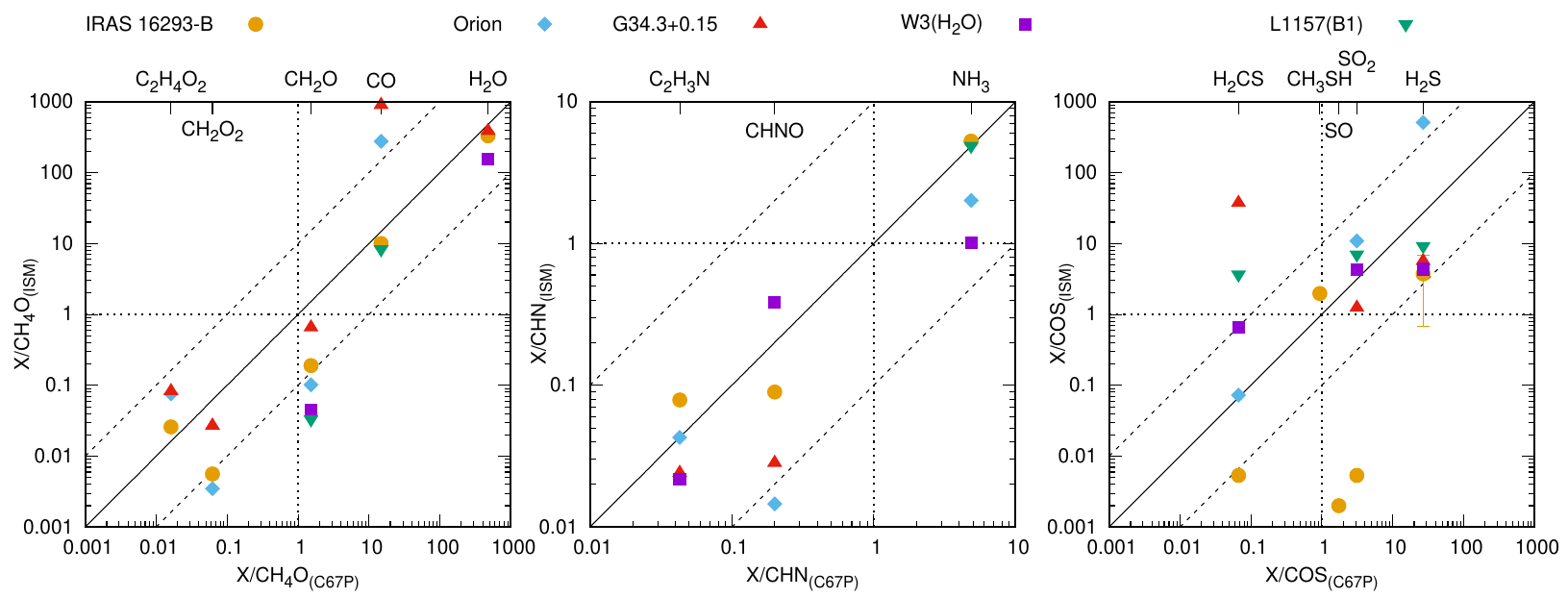}
\caption{\label{fig:ISM67P}Relative abundances in comet 67P (x-axis) and several objects in the ISM (y-axis). Left: O-bearing species, normalized to methanol, CH$_4$O. Middle: N-bearing species normalized to hydrogen cyanide, CHN. Right: S-bearing compounds normalized to carbonyl sulfide, COS. A perfect correlation would be found along the black solid lines and within a factor 10 between the the dashed lines. Data were taken from \cite{Rubin2019a,Bockelee2000,Drozdovskaya2019} and isomers were added up, if they could not be distinguished by ROSINA.}
\end{figure}

Complex organic molecules (COMs) commonly refer to species with at least six atoms including one or more carbon atoms \cite{Herbst2009}. COMs are detected in numerous astrophysical environments and are a key field of research [C. Ceccarelli et al. in this volume]. Also comets are known to host a vast number of organic and inorganic molecules \cite{DelloRusso2016a,Biver2019a} and comet 67P is no exception \cite{Altwegg2019,Rubin2019b}. Numerous C-H-O-N-S-bearing molecules have been identified including the most simple amino acid, glycine \cite{Altwegg2016,Elsila2009} as well as a substantial amount of saturated and unsaturated organic molecules. Based on the census of molecules found in the coma of 67P, several conclusions have been derived: similar to comet \mbox{C/1995 O1 (Hale-Bopp)} \cite{Bockelee2000}, also the relative abundances in comet 67P  \cite{Drozdovskaya2019} exhibit comparable relative abundances as observed in the ISM (see Figure~\ref{fig:ISM67P}). These findings highlight the similarity of cometary and interstellar ices, possibly hinting at a formation under similar physical and chemical conditions.

Furthermore, the lack of nitrogen in cometary material, which was already observed at 1P/Halley \cite{Geiss1988}, has been confirmed for the gas phase of 67P \cite{Rubin2019a}. Nitrogen being locked-up in ammonium salts may, however, explain the lack of N in the gas phase \cite[ O. Poch et al. in this volume]{Altwegg2020a,Poch2020}. Similarly, substantial amounts of sulfur may also be locked-up in salts \cite{Altwegg2022} or more refractory S-bearing polymers up to S$_8$ \cite{Calmonte2016}, both challenging for remote observations. In turn this may then account for the reported lack of gas phase sulfur in dense clouds and star-forming regions \cite{Penzias1971,Tieftrunk1994}.

\subsection{\label{sec:d2h}D/H in cometary water}
A key measurement of ROSINA was D/H in cometary H$_2$O (see Tab.~\ref{tab:isotopes}). The D/H ratio in comets ranges from terrestrial to several times this ratio \cite{Bockelee2015a,Lis2019}. This variability suggests that comets formed over a large range in heliocentric distances and that different families of comets (JFCs, OOCs, etc.) did not form separately, but rather have a  different dynamical history. The D/H ratio is also related to the question whether the water on Earth could be of cometary origin. The D/H ratio of singly deuterated water $\left(\frac{D}{H}=\frac{1}{2}\frac{HDO}{H_2O}\right)$ in comet 67P, however, was more than three times the ratio measured on Earth \cite{Altwegg2015}. Furthermore, the D/H of doubly deuterated water $\left(\frac{D}{H}=2\frac{D_2O}{HDO}\right)$ shows an enhancement of more than a factor two hundred \cite{Altwegg2017a} compared to the terrestrial ratio. This then implies that comets, such as 67P, cannot be a major contributor to the water on Earth. 

The different isotope ratios in single and doubly deuterated water also have implications for the formation of comet 67P: such a discrepancy has already been observed in the ISM, e.g., at low-mass protostar NGC 1333 IRAS2A \cite{Coutens2014}, and may be explained by the chemical evolution in the early cold stages of the formation of molecular clouds to protostellar cores \cite{Furuya2016}. This implies that the H$_2$O in 67P was formed well before the comet and temperatures below $\sim$70~K throughout the lifetime of the comet prevented H-D isotope exchange reactions \cite{Lamberts2015}.

On the other hand, the measured D/H ratio did not differ among the two lobes of 67P \cite{Schroeder2019b}, which is consistent with the cometesimals involved in the collisional merger \cite{Jutzi2015} originating from the same region or similar heliocentric distance in the protoplanetary disc. \cite{Mueller2022} furthermore showed that, within errorbars, the D/H did not vary during the apparition, despite the comet's outgassing activity changing by several orders of magnitude through perihelion \cite{Combi2020}.

\subsection{\label{sec:isotopevariations}Isotope variations}
Aside from D/H in water, ROSINA measurements in the coma of comet 67P yielded isotopic ratios of several elements. Table~\ref{tab:isotopes} lists a collection of isotope ratios in C-H-O-N-S-bearing molecules. The corresponding protosolar ratios have been added for reference and a color scheme has been adopted to highlight substantial deviations. Red indicates elevated relative abundances of the heavier, more rare isotope such as D (versus H) or $^{18}$O (versus $^{16}$O) compared to the protosolar ratio. Green indicates a similar relative isotope abundance, i.e. within more or less 1-$\sigma$ uncertainty, while blue marks lower abundances of the heavier, more rare isotope. Additionally, deviations from protosolar isotope ratios have been measured in silicon \cite{Rubin2017} and the noble gas xenon \cite{Marty2017}, while the ratios in chlorine and bromine \cite{Dhooghe2017} and the noble gases argon and krypton \cite{Rubin2018} were consistent within errorbars with protosolar relative abundances (see \cite{Hoppe2018} for a review). In summary, the marked deviations in isotopic ratios indicates non-homogeneous mixing of the material at the location where comet 67P formed in the early solar system.

\begin{table}[ht]
\caption{\label{tab:isotopes}Isotope ratios in comet 67P measured by ROSINA \cite{Altwegg2019,Altwegg2017a,Mueller2022,Hoppe2018,Calmonte2017,Altwegg2020b,Hassig2017,Schroeder2019a,Drozdovskaya2021}. Colors refer to the increased (red), comparable (green), or lower (blue) heavy isotope abundances with respect to the corresponding protosolar ratio \cite{McKeegan2011,Marty2011,Lodders2010}. A subset of the sulfur isotopes were measured in multiple time periods during the Rosetta mission. \vspace{0.05cm}}

\resizebox{\textwidth}{!}{
\begin{tabular}{lcccccccl}
  Species & D/H & $^{12}$C/$^{13}$C & $^{14}$N/$^{15}$N & $^{16}$O/$^{17}$O & $^{16}$O/$^{18}$O & $^{32}$S/$^{33}$S & $^{32}$S/$^{34}$S & notes\\
  \hline
   & 1.94$\cdot$10$^{-5}$ & 98$\pm$2 & 441$\pm$6 & 2798 & 530 & 126.7 & 22.5 & protosolar reference \\ \hline \hline
   
   H$_2$O & \cellcolor{red!25}\begin{tabular}{c} (5.0$\pm$0.4)$\cdot$10$^{-4}$ \\ (3.6$\pm$1.8)$\cdot$10$^{-2}$ \end{tabular}  & & & \cellcolor{red!25}2347$\pm$191 & \cellcolor{red!25}445$\pm$45 &  & & \begin{tabular}{c} D/H in HDO/H$_2$O \\ D/H in D$_2$O/HDO \end{tabular} \\ \hline

  CO$_2$ &  & \cellcolor{red!25}84$\pm$4 &  & & \cellcolor{red!25}494$\pm$8 & & & \\ \hline

  CO &  & \cellcolor{green!25}86$\pm$9 & &  &  &  &  & \\ \hline

  CH$_4$ &  \cellcolor{red!25}(2.41$\pm$0.29)$\cdot$10$^{-3}$  & \cellcolor{green!25}88$\pm$10 & & & & & & \\ \hline

  C$_2$H$_6$ &  \cellcolor{red!25}(2.37$\pm$0.27)$\cdot$10$^{-3}$  & \cellcolor{green!25}93$\pm$10 & & & & & & \\ \hline

  C$_3$H$_8$ &  \cellcolor{red!25}(2.16$\pm$0.26)$\cdot$10$^{-3}$  & \cellcolor{green!25}87$\pm$9 & & & & & & \\ \hline

  C$_4$H$_{10}$ &  \cellcolor{red!25}(2.05$\pm$0.38)$\cdot$10$^{-3}$  & \cellcolor{green!25}96$\pm$14 & & & & & & \\ \hline

  H$_2$CO &  & \cellcolor{red!25}40$\pm$14 &  & & \cellcolor{red!25}256$\pm$100 & & & \\ \hline

  CH$_3$OH &  \cellcolor{red!25}(0.71 to 6.6)$\cdot$10$^{-2}$  & \cellcolor{green!25}91$\pm$10 &  & & \cellcolor{green!25}495$\pm$40 & & & \\ \hline

  O$_2$ & & & & \cellcolor{red!25}1544$\pm$308 & \cellcolor{red!25}345$\pm$40 &  & & \\ \hline

  NH$_3$ & \cellcolor{red!25}(1.1$\pm$0.2)$\cdot$10$^{-3}$ & & \cellcolor{red!25}118$\pm$25 & & & & & \\ \hline

  NO & & & \cellcolor{red!25}120$\pm$25 & & & & & \\ \hline

  N$_2$ & & & \cellcolor{red!25}130$\pm$30 & & & & & \\ \hline

  H$_2$S & \cellcolor{red!25}(1.2$\pm$0.3)$\cdot$10$^{-3}$ & & & & & \cellcolor{blue!25}\begin{tabular}{c} 187$\pm$9 \\ 132$\pm$3 \end{tabular} & \begin{tabular}{c} \cellcolor{blue!25}23.6$\pm$0.4 \\ \cellcolor{green!25}22.4$\pm$1.6 \end{tabular} & \begin{tabular}{l} $^{32}$S/$^i$S: Oct 2014 \\ $^{32}$S/$^i$S: May 2016 \end{tabular}\\ \hline
 
  SO & & & & & \cellcolor{red!25}239$\pm$52 &  & \begin{tabular}{c} \cellcolor{green!25}23.5$\pm$2.5 \end{tabular}& \\ \hline

  SO$_2$ & & & & & \cellcolor{red!25}248$\pm$88 &  & \begin{tabular}{c} \cellcolor{green!25}21.3$\pm$2.1\end{tabular} & \\ \hline

  OCS & & & & & \cellcolor{red!25}277$\pm$10 & \begin{tabular}{c}  \\  \\ \cellcolor{blue!25}165$\pm$12 \end{tabular} & \begin{tabular}{c} \cellcolor{blue!25}25.1$\pm$1.3 \\ \cellcolor{green!25}21.7$\pm$4.0 \\ \cellcolor{green!25}22.8$\pm$0.3 \end{tabular} & \begin{tabular}{l} $^{32}$S/$^i$S: Oct 2014 \\ $^{32}$S/$^i$S: Mar 2016 \\ $^{32}$S/$^i$S: May 2016 \end{tabular} \\ \hline

  CS$_2$ & & & & & & \cellcolor{blue!25}\begin{tabular}{c} 157$\pm$7 \\ 151$\pm$8 \end{tabular} &\begin{tabular}{c}  \cellcolor{blue!25}24.3$\pm$0.7 \\  \cellcolor{blue!25}25.3$\pm$0.6 \end{tabular} & \begin{tabular}{l} $^{32}$S/$^i$S: Oct 2014 \\ $^{32}$S/$^i$S: May 2016 \end{tabular}\\

\end{tabular}
}
\end{table}

\subsection{\label{sec:noblegases}Transport of cometary material to the early Earth}
The xenon in 67P shows a marked depletion in the two heavy isotopes $^{134}$Xe and $^{136}$Xe \cite{Marty2017} and thus bears some resemblance to the previously postulated primordial Xe, or so-called U-Xe \cite{Pepin2000}, required to explain the peculiar Xe isotope ratio in the terrestrial atmosphere. \cite{Marty2017} thus estimated that some 22~$\pm$~5\% of the xenon in today's terrestrial atmosphere is of cometary origin. Taking Xe and its relative abundance with respect to other cometary  species as baseline, the transport of cometary material to the early Earth can be estimated \cite{Rubin2019a}. As a result, and in agreement with section~\ref{sec:d2h}, less than 1\% of the water on Earth would hence be of cometary origin \cite{Marty2016,Rubin2019b}, however, the contribution of organic molecules may be substantial when compared to the Earth's biomass \cite{Rubin2019a}.

\section{Summary and conclusions}
Several key measurements of ROSINA indicate that the ices in 67P are inherited from stages preceding the formation of our solar system \cite{Altwegg2019}: relative abundances of the molecules measured in the coma of the comet resemble those measured in the ISM \cite{Bockelee2000,Drozdovskaya2019}. Furthermore the D/H ratio in single versus double-deuterated water, abundant amounts of O$_2$, and the presence of highly volatile species hint at limited processing of the cometary material since incorporation \cite{Altwegg2017a,Furuya2016,Rubin2020}. Pronounced differences in the isotopic ratios of a suite of volatile species compared to the protosolar ratios also suggest that mixing of the material in the solar system prior to the incorporation into the comet was rather limited \cite{Altwegg2019,Rubin2020}. Also, comets such as 67P, may have contributed substantially to the terrestrial atmosphere \cite{Marty2017} and prebiotic organic inventory \cite{Rubin2019b} but not so much to the terrestrial oceans \cite{Marty2016,Rubin2019b}.

\paragraph{\bf Acknowledgements}
The work of the many engineers, technicians and scientists involved in the Rosetta mission and in the ROSINA instrument in particular is gratefully acknowledged. Without their contributions, ROSINA would not have produced such outstanding results. Rosetta was an ESA mission with contributions from its member states and NASA. Work on this paper at the University of Bern was funded by the Canton of Bern and the Swiss National Science Foundation (200020\_207312).


%
%

\end{document}